# Bug Fix Time Optimization Using Matrix Factorization and Iterative Gale-Shaply Algorithms

Madonna Mayez
Software engineering
*British University in Egypt*
Egypt
Madonna.mayez@bue.edu.eg

Khaled Nagaty
Computer Science
*British University in Egypt*
Egypt
Khaled.nagaty@bue.edu.eg

Abeer Hamdy
Software engineering
*British University in Egypt*
Egypt
Abeer.hamdy@bue.edu.eg

*Abstract*— Bug triage is an essential task in software maintenance phase. It assigns developers (fixers) to bug reports to fix them. This process is performed manually by a triager, who analyzes developers' profiles and submitted bug reports to make suitable assignments. Bug triaging process is time consuming thus automating this process is essential to improve the quality of software. Previous work addressed triaging problem either as an information retrieval or classification problem. This paper tackles this problem as a resource allocation problem, that aims at the best assignments of developers to bug reports, that reduces the total fixing time of the newly submitted bug reports, in addition to the even distribution of bug reports over developers. In this paper, a combination of matrix factorization and Gale-Shapely algorithm, supported by the differential evolution is firstly introduced to optimize the total fix time and normalize developers work load. Matrix factorization is used to establish a recommendation system for Gale-Shapley to make assignment decisions. Differential evolution provides the best set of weights to build developers score profiles. The proposed approach is assessed over three repositories, Linux, Apache and Eclipse. Experimental results show that the proposed approach reduces the bug fixing time, in comparison to the manual triage, by 80.67%, 23.61% and 60.22% over Linux, Eclipse and Apache respectively. Moreover, the workload for the developers is uniform.

## I. Introduction

During the maintenance phase of software projects, bugs are reported, with different levels of severity. Some of them require immediate solving with high level of complexity, whereas others are minor changes, which have no serious effect on the software under test and need low level of experience to be fixed. Bug reporters and fixers use issue tracking systems (ITS) to communicate, report, assign and propose solutions for the raised issues. Issue tracking systems such as Bugzilla [1] and Jira [2] enable the developers to track the massive amount of daily change requests.

Bug reporter uses bug report standard document to submit a change about the software under test. It has some fields that give some details about the bug, such as bug ID, summary, description, reporter, assignee, open date; when the developer starts working on that bug and closed date; when the bug is totally solved and closed. Bug severity level shows the bug report effects on the overall software. Bug priority level shows the urgency level of solving this bug because there may be some other bugs are depending on solving this bug. In addition, bug reporter and fixers and communicate through some comments discussing unclear aspects on the bug report.

Starting from submitting a bug report on ITS, it goes under some sequential phases holding different status in each. NEW status is the first one after submitting the bug report. It changes to be ASSIGNED, if a developer is assigned to that bug report. INVALID status means that there is no solution for that bug. CLOSED status means that a bug report is either fixed or cannot be fixed. After a managerial revision for the bug, it should be marked as REVISED. If there is another report discussing the same problem, it holds DUPLICATE status.

Bug triaging task, is the manual process of assigning developers to the submitted bug reports for fixing the bugs [3]. As a real-life scenario, the team leader studies the existing developers' profiles and the newly submitted bug reports and match developers' qualifications with the experience level required in each bug report.

Triaging process could be manageable with a small number of bugs reports and developers. However, statistics opposes that. For example, more than 500,000 report are submitted on Eclipse ITS in 2019, and in 2012, Mozilla received above 800,000 reports [4]. These numbers indicate the complexity of bug triaging process. Because it is manually executed, there must be some mistakes that lead to bug tossing.

Tossing a bug means reassign it to another fixer because the originally assigned developer cannot solve it due to the lack of experience. More than 37% of the assigned bug reports are tossed [5]. These sort of practices not only waste the financial and human resources, but also it delays the fixing time and delivering the whole software. Optimal assignment for a bug report is required because over loaded developer takes much time to address the assigned bug reports and less qualified developer takes much time trying to solve the assigned bug, but he cannot. Therefore, optimal triaging cannot be manual because of the human mistakes. There must be an automated system, that matches between bug reports and developer sets considering the developer experience required for a bug report and qualifications level of developers.

Numerous studies have been proposed to automate the bug triaging problem, some studies are based on information retrieval approach [6] [7] [8] [9] [10], others used machine learning classification techniques [11] [12] [13] [14] [15] [5]. Some research studies considered the developer workload, which has a huge impact on the total fixing time. However, the optimization percentage is relatively low in most of them. This paper aims at establishing a triaging system that enhances the fixing time optimization percentage besides equalizing the developers' workload.

*A. Aims and contribution*

This paper proposes an approach for automating the bug triaging task such that, the bug fixing time is reduced and the developer workload is normalized. The contributions of this work can be summarized as follows:

- The triaging task is formulated as a task-resource allocation problem. Gale-Shapely algorithm is used to find the optimal assignment of the developers to the newly submitted bug reports.

- In order to apply Gale-Shapely algorithm, a score is calculated for every developer-bug report pair (developer score) that expresses the developer's experience in fixing that bug report.

- Topic modelling is used to categorize and label the bug reports, in addition to determining the areas of experience of the different developers.

- Developer score is calculated by three bug report attributes. To set a weight for each factor, ddifferential evolution algorithm is applied, with restriction on the summation of the three weights to equal to 1 and each weight must be greater than 0.

- The two inputs matrices to Gale-Shaply algorithm (developers and bug reports) should be equal in size, however, this is not the real case, thus Gale-Shapley algorithm is applied iteratively, to ensure the one-to-one assignment, that leads to the normalized workload.

Paper Organization: this paper is organized as follows. Section II summarizes the previous related work for automating bug triage with the related techniques and the research gap in the existing bug triage approaches. Section III introduces the proposed model including the fundamental concepts, experiments and the evaluation matrices to measure the effectiveness of the proposed model. Section IV reports the experimental results and analysis. Section V is the treats of validity for the proposed model. Finally, section VI is a conclusion and some future work for further extensions.

## II. LITERATURE SURVEY

Several studies were proposed to automate the bug triaging process. Machine learning and information retrieval are the most common approaches. Graph theory concept also used by few researchers. This section gives a brief summary for the state of art for handling bug triaging problem.

*A. Information retrieval techniques*

Sahu et al [7] followed the feature selection approach to reduce the dataset size. A combination of K-Nearest Neighbor and Naive Bayes approaches (KNN+NB) are used to extract the features from the dataset. Sahu et al used Mozilla bug repository to evaluate the proposed approach. Prediction accuracy recorded is 85%. Sun et al [9] also reduced the dataset size by detecting the duplicate bug reports. They assumed that filtering the dataset from duplicate bug reports will reduce open bug reports and accordingly the fixing time will be optimized, because instead of two fixers are busy by the same bug, one will be in charge. Sun et al bug duplicate detector is called REP, it extends the BM25F similarity matrix. Bug reports from Mozilla [16], Eclipse [17] and OpenOffice [18] are extracted to evaluate the prediction accuracy, which ranges between 37 and 71% in recall and 47% in mean average precision.

Nguyen et al [10] followed Sun's strategy in optimizing the triaging code by identifying the duplicated bug reports and avoiding them. Their proposed approach, duplicate bug report detection approach (DBTM), is an extension for Sun's approach. It is a combination of information retrieval-based features and topic-based features. Each record is represented as a textual document describing the technical issue in the system. Duplicated records are detected by textual similarity method. For feature extraction, IR and topic modeling are applied using ensemble averaging technique, the mean of quantity. Markov chain Monte Carlo (MCMC) algorithm is used as a sampling technique to record the observation in a sequential approach. As a result, DBTM reduced the training dataset by more than 20% in comparison with the previous work.

XIA et al [8] used topic modelling approach in their proposed solution, called TopicMinerMTM. It is an extension for Latent Dirichlet Allocation (LDA) algorithm [19]. The distribution of the topic modeling approach is used to calculate the degree of matching between developers and bug reports. Their approach is evaluated over GCC [20], NetBeans [21], Eclipse [17], OpenOffice [18], Mozilla [16]. In total, 227,278 instance of bug reports are used. The achieved accuracy ranges between 48% and 90%. Park [22] also used LDA to categorize bug reports for the purpose of enhancing the prediction accuracy in their proposed solution, COSTRIAGE. To formulate a developers' profile Arun's is used to determine the optimal number of K and accordingly average fixing time is used to represent the developer experience with respect to the topic modeling results. To evaluate COSTRIAGE, bug reports from Linux, Apache, Mozilla and Eclipse bug repositories are extracted. Using 80% of the dataset for training the model and 20% for testing, results show that fixing time is reduced by more than 30% in Apache dataset, with 4.5% to 69.7% in prediction accuracy.

*B. Machine learning techniques*

ML algorithms are used to handle bug triaging process as a classification problem. To train a classifier, historical bugs are leveraged and considering that each developer is a class, the prediction step is executed to assign bug reports in testing dataset to the existing developers set.

Kashiwa et al [11] are the first to focus on the developer's workload factor. They believe that distributing the bug reports equally on the developers set will optimize the total fixing time. They formulated the triaging problem as a knapsack problem. They aim to find the optimal assignment between bug reports set and developers set, considering the developer workload. SVM classifier and Laten Dirichlet allocation are used for developer prediction. Mozilla Firefox [16], Eclipse [17] Platform and GNU compiler collection (GCC) [20] are used to evaluate the proposed approach. The total fixing time is reduced by 35%-41%, in comparison with the manual triaging.

S. Mani et al [13] proposed a deep learning-based model called, deep bidirectional recurrent neural network (DBRNN-A). Syntactic and semantic features are considered by long word sequence. Description and title are the main focus in each bug report. multinomial naive Bayes, cosine distance, support vector machines, and SoftMax are used to predict the

1. NLTK :: Natural Language Toolkit

optimal assignment in a comparative approach. To evaluate the proposed solution, 383,104 bug report from Google Chromium, 314,388 bug report from Mozilla Core and 162,307 bug reports from Mozilla Firefox are exported. Results show that DBRNN-A along with Softmax is better than the bag of words model.

Bhattacharya et al [14] used bag of words and frequency-inverse document frequency (TF-IDF) techniques for feature extraction. Title, description, keywords, products, components, last developer activity are the extracted bug fields in each record. TF-IDF is used to simulate the importance of word in each document. Naive Bayes, Bayesian network and tossing graph are used as classifiers. The dataset used is quite large, 306,297 bug reports from Eclipse [17] and 549,962 from Mozilla [16]. Experimental results showed that the proposed solution achieved 83.62% prediction accuracy. By considering the report ID, the average prediction accuracy became 77.64%. However, the prediction accuracy dropped to 63% while ignoring the bug report ID.

Gondaliya et al [5] used text mining methods. First, lemmatization is applied to reduce the inflectional forms. Second, part-of-speech tagger (POS tagger), which is a software for reading a text in different language. It is used to assign a part of speech for each term. Third, bigram method is applied to extract the probability of tokens given another token. For the purpose of classification, Linear Support Vector Machines (SVMs), multinomial naive Bayes, and Long Short Term Memory (LSTM) networks are applied. In small dataset (1215 bug report), the prediction accuracy ranges between 47.9 and 57.2%. For the large dataset it ranges from 68.6% → 77.6%.

*C. Graph theory*

Zaidi et al [23] focused-on graph representation theory to handle the triaging problem. A heterogeneous graph is built using description and summary fields from each bug report. The first stage in the proposed solution simulates word-to-word co-occurrence graph and the second one is word-to-bug report. To weight the graph, point-wise mutual information (PMI) is applied in the first stage. TF-IDF is used to represent the word co-occurrence in the second stage. Softmax algorithm is used to classify the new bug report and assign a developer for each record. Five different datasets from Bugzilla and Firefox are used to evaluate the proposed approach. Among many distance measurement approaches used, such as PMI, cosine, Jaccard index, and Euclidean distance, PMI recorded the best performance. Comparing with the previous automating triaging approaches, predicting top-1 accuracy gets higher by 3% to 6% and up to 5% to 8% in top-10 accuracy.

Yadav et al [24] came up with a three-stage novel strategy called developer expertise score DES. First, an offline process assigns a score for each developer based on versatility, priority and the average fixing time. Finding the capable developers is handled by applying the similarity measures such as namely feature-based, cosine-based similarity and Jaccard. Second, ranking the developers according to their DES. Finally, Navies Bayes, Support Vector Machines and C4.5 classifiers are used for classification purpose and they compared their performance with the ML-based bug triaging approaches. By evaluating the algorithm over five open-source datasets (Mozilla [16] ,Eclipse [17], Netbeans [21], Firefox [16]) with 41,622 bug reports, results showed that DES systems recorded 89.49% in mean accuracy, 89.53% in precision, 89.42% in recall and 89.49% in F-score. Additionally, the bug tossing length is reduced by 88.55%.

*D. Differences from previous work*

The previous researchers handled the triaging problem as an information retrieval or classification problem, considering the manual triaging as the optimal one. Additionally, distributing the workload over the developers in a uniform way was disregarded by the majority of the previous work. This paper focuses on optimizing the total fixing time of the submitted bug reports, in addition to the even distribution of the workload over developers.

The work proposed by Park et al aims at optimizing the fixing time same as this work. However, Park et al models the developer profile using only the fixing time factor. This is not enough to simulate the developer skills. The developer fixing time could be affected by the developer's experience in fixing sever bugs and working on different bug components. For example, if a developer used to fix sever bugs, it is expected to fix new bugs with similar level of severity in shorter time. In addition, if a developer is assigned to Proxy bug for the first time, he will spend much more hours/ days compared with another developer who is familiar with bugs of Proxy component. Park, used 80% of the dataset in training and 20% in testing. This is not an accurate data split for a sort of problems that consider time like triaging problem.

Previous researchers handled the triaging problem as an information retrieval or classification problem. Thus, most of them took similar stages to solve the problem using different algorithms. Taking the manual triaging as the optimal one, XIA et al [9] achieved the highest prediction accuracy percentage, near to 90%. Additionally, distributing the workload over the developers in a uniform way is ignored by most of researchers. This work proves that this concept has a significate effect on optimizing the total fixing time. Researchers, who took the tossing graph reduction approach, reduced the TGL and improved the triaging accuracy compared with the manual triage. However, they ignored the workload distribution for the developers. As a result, small group of developers will be overloaded and accordingly, the bug fixing time will increase.

Park's model is the closest model to this paper. Its main goal is the fixing time optimization. However, Park et al models the developer profile using only the fixing time factor. This is not enough to simulate the developer skills. The developer fixing time could be affected by the developer's experience in fixing sever bugs and working on different bug components. For example, if a developer used to fix sever bugs, it is expected to fix new bugs with similar level of severity in shorter time. In addition, if a developer is assigned to Proxy bug for the first time, he will spend much more hours/ days compared with another developer who is familiar with bugs of Proxy component. Park, used 80% of the dataset in training and 20% in testing. This is not an accurate data split for a sort of problems that consider time like triaging problem.

Different from Park, this paper calculates the developer score for recording the developer experience level using three factors, severity level, component and fixing time. Each factor has an effect weight on calculating the developer score, which is optimized by the differential evolution algorithm, seeking the highest fixing time reduction. In addition, time split approach is used to evaluate the proposed solution. Dataset records are put in chronological order and just a part of dataset is considered and it is gradually increased in each iteration, starting form 20% in the first iteration reaching 100% in the 9th iteration. In each step, the latest 10% is considered the testing dataset (new submitted bug reports) and the remaining first part is the training dataset (historical bug reports). Moreover, the proposed solution considers the developer workload, which is ignored in most of research papers including Park's work.

---

1. NLTK :: Natural Language Toolkit
2. glove · PyPI

This work proposes an optimization approach for the bug triaging process using three different datasets with different sizes, Linux, Eclipse and Apache. It presents the effect of the Gale-Shapely algorithm, supported by matrix factorization and Differential Evolution algorithm, on optimizing the fixing time and shows its advantage in normalizing the developer workload.

### III. PROPOSED APPROACH

The proposed solution is built with intention to reduce the bug fixing time and automate the triaging problem by normalizing the developer workload. In this way, all developers will be loaded by equal number of bug reports, which means, all of them will work in parallel to fix the submitted bug reports. Instead of the manual triaging which loads only a group of developers with a big number of bugs and other group with zero load.

The proposed approach steps can be summarized as followed:

- As each bug report must be assigned to a developer, this paper formulates the triaging problem as a task-resource allocation problem. Gale-Shapely algorithm is used to find the optimal assignment between bug reports and developers. The input of Gale-Shapely algorithm is two numerical matrices, the first one indicates how the skills required in each bug report matches the developers experience and the second one is about how each developer is skilled in the fields of the new bug reports. The first matrix can be considered as a bipartite graph. The two sets, bug reports and developers, form a complete bipartite graph because each node in S1= {BR1, BR2, BR3…} has a relation (weighted edge) to be linked with each node in S2= {D1, D2, D3…}. That link simulates the developer experience (developer score) in the corresponding bug report.

- For the first matrix, in order to perform this calculation (scoring phase), topic modeling technique is applied to work as a labeling stage for the purpose of differentiating between the fields of the bug reports. The topics (labels) of the bug reports also represent the skills of the developer. The result of this step is that each bug report will be labeled with the topic that contributes with the maximum weight. Therefore, a score per topic will represent each developer. Thus, a vector of scores simulates the skills of each developer in each topic. Each number in the vector is calculated by the history of the developer, which is represented by three main factors; the severity, component and the median fixing time that the developer took to fix bug reports of topic K.

- The three factors used to calculate the developer score contribute with different weights. The differential evolution is used to set weight optimization stage; with restriction on the summation of the three weights must equal to 1 and each weight must be greater than 0.

- For the second matrix, it works as a recommending system. It is the result of multiplying the score matrix by the topic modelling result matrix. Further explanation will be provided in the following section.

- Gale-Shapely algorithm is used to achieve one to one assignment using the two previously mentioned matrices to assign a developer to each bug report. To ensure that each bug report is assigned to a developer, the number of elements in the two matrices should be equal. However, the number of bug reports is much bigger than the number of developers. For this reason, this paper applies Gale-Shapely iteratively. The bug report set will be divided into small subsets, each of them will be of size that equals to the size of the developers set. In each iteration, the developers set will be distributed equally on the set of bug reports considering the bug report type and the developer score. As a result, the workload will be normalized across all developers. This process will be repeated in a loop with length that equals to the testing dataset size (newly submitted bug reports) divided by the size of the developers set.

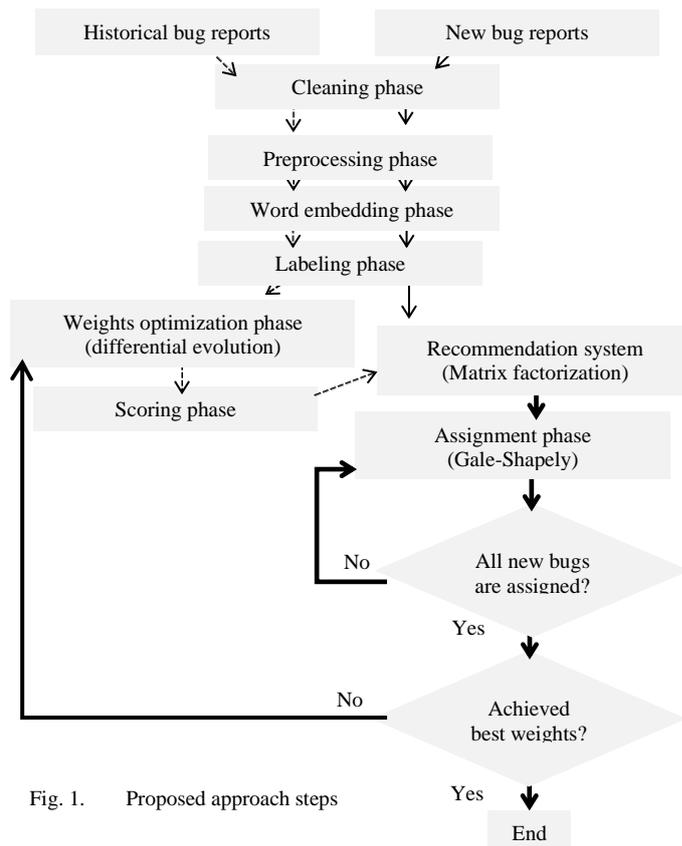

Fig. 1. Proposed approach steps

Fig 1 depicts the steps of the proposed approach, which are discussed in details in the following subsections.

At the beginning, a text cleaning and reduction techniques will be applied. For example, only bug report of "CLOSED" and "RESOLVED" status are considered and the records with developers who handled less than 10 bug reports are removed, similar to what Xuan et al. [25] applied in his work. After that, each bug report goes under some preprocessing techniques, such as tokenization, stop word removal, lemmatization and word embedding. A labeling process will be applied using the topic modeling technique. Using the labeling step results and the developer history, a score is assigned to each developer. For representing the bug reports and the developers, a bipartite graph was used for modeling the two sets as vertices. Third, a recommendation system is applied using matrix factorization between the matrix of the developer scoring (row: developer, col: topics) and the matrix of the topic modeling result (row: topics, col: bug reports). Gale-Shapely algorithm will be applied in an iterative manner on the result of the recommendation system and bipartite graph to get one-to-one assignment between bug reports and developers.

---

1. NLTK :: Natural Language Toolkit

2. glove · PyPI

*A. Fundamental concepts:*

*1) Cleaning phase:*

In this phase, the dataset is cleaned from any useful records. For example, records with empty cells in any of the used bug report fields (description, assignee, status, open date, closed date, severity and component) will be removed from the dataset. On top of that, records with assigned developer, who fixed less than 10 records across all the dataset, will be removed. To ensure that each record has an assigned developer (the fixer), only records with status "CLOSED" and "RESOLVED" are considered.

*2) Preprocessing phase:*

The textual part (description field in each bug report) goes under some preprocessing algorithms using natural language processing (NLP) techniques

Tokenization: is the first step in preprocessing the textual data. It works on splitting the corpus into small parts, whether to be a word or sentence. The algorithm is implemented using an open-source library called NLTK1 (Natural Language Toolkit). It can handle more than one tokenizer such as word and sentence tokenizer, Treebank word tokenizer and punctuation-based tokenizer [26].

Stop word removal: this technique is implemented for the purpose of removing the words such as "a", "an", "who", "that", "the" and "in" that make some noise in the text, which defiantly affect the processing time. These terms are considered to be not very discriminatory. In other words, their effect in retrieving information or classification is not counted. NLTK library is also used to apply this technique [27].

Lemmatization: this approach is responsible for removing the inflectional ending in each term. Thus, it returns the terms in each document in dictionary form, considering the term type whether to be verb or noun. Additionally, it matches each term with its synonyms. For example, when searching for the word "hot", the term "warm" also matches [28].

*3) Word embedding phase*

In real life, the bug description is written by developers (human), who can use different terms to explain the same concept. So, Glove technique is implemented to normalize the words across the whole corpus (bug reports represented by description fields). It represents the words of the same meaning with similar representation. In this way, the semantics will be precisely handled.

GloVe[2] (Global Vectors for word Representation): used for word embedding. It is unsupervised method which allows words with similar semantics and meaning to have similar representation. It is considered as an extension for word2vec technique. It is developed by Pennington, et al [29], which provides a vector representation of each word. Basically, it combines the global factorization techniques such as LSA (Latent Semantic Analysis) with Word2vec: the local context-based technique. So, it uses statistics to represent the word co-occurrence throughout the whole corpus.

*4) Labeling phase:*

In this step, each bug report will be titled by a specific label, which also represents the field of skills needed for the assigned developer. Latent Dirichlet allocation (LDA) [30] is a topic modeling approach, that can be used for this purpose. The bug report will be represented as a vector of percentages, simulates to what extent this bug report is related to each topic. Among this group of percentages, the bug report will be labeled by the topic with maximum weight.

Topic modeling: an unsupervised technique, that used for the purpose of extracting the main topics from the corpus (description field of the historical bug reports). Each topic is represented by a set of words with different probabilities. The bug reports are simulated by vectors, each of which represent the occurrence probabilities of the topic in the bug report. LDA technique is used to apply the labeling phase.

A graphical representation for the LDA is shown in Fig 2, α is the density of the document in each topic, β is topic-word density, the number of documents across the whole corpus is represented by M. In each document, N represents the number of words, while θ is topic distribution for document M. Each word w is assigned to a specific topic z. Finally, w is the word in α topic.

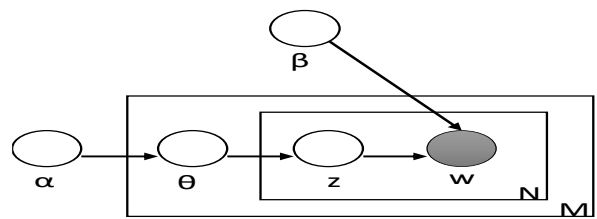

Fig. 1. Graphical representation for LDA [8]

Hybrid attributes (α and β) depends on the corpus documents. Ideal model requires accurate values for these two parameters. Large values for α denotes more topics in each document and vise-versa. On the other hand, large values for β represent that each topic has more words and vise-versa.

Choosing the number of topics (K)

To execute LDA algorithm, the number of topics needs to be specified. A part of code is written in order to calculate coherence score for a number of topics that is ranged from 1 to K. This score simulates to what extent similar words of the same topic are logically related. This process is done thought providing probabilistic coherence score for each number. The higher the value of K, the more sense words related to each other in each topic and accordingly, more expressive the topic will be.

*5) Scoring phase*

The goal of this phase is to formulate the developer experience in form of numbers. The training dataset (historical bugs) is used to get the history of each developer. In each bug report, the median fixing time, bug severity and bug components are used to calculate the developer score. The experience of each developer will be represented as a vector of scores (score per topic). Thus, the score vector length is equal to the optimal number of K. Considering that BR is a bug report. $BR_D^K$ in Equation 1 is a set of bug report of topic K solved by developer D. The following equations simulates the steps of calculating the developer score.

---

1. NLTK :: Natural Language Toolkit

2. glove · PyPI

$$BR_D^K = BRs\ of\ topic\ K\ solved\ by\ developer\ D \quad (1)$$

Bug severity:

It is believed that the developer who can fix sever bug reports is more experienced than who fix bug report with low severity. So, the higher the bug report severity is, the more experienced the developer will be. Bug severity levels are critical, normal, minor, trivial and traditional. Equation 2 calculates the severity score for each developer considering the eight severity levels. $w_l$ is a weight used to differentiate between the levels $S_l$, with values (0.29, 0.21, 0.16, 0.11, 0.09, 0.07, 0.05, 0.02), for levels blocker, critical, major, normal, minor, trivial, regression and enhancement, as an assumption.

$$S(D)_{sev} = \frac{\sum_{l=1}^{l=8} w_l * (BR_D^K | S_l)}{\sum BR_D^K} \quad (2)$$

Bug component:

In fact, the more component types in $BR_D^K$, the more experienced the developer will be, as the developer will be aware by many aspects in the bug report. Equation 3 represents how developer component score $S(D)_{com}$ is calculated.

$$S(D)_{com} = \frac{\#component\ types\ in\ BR_D^K}{all\ component\ types} \quad (3)$$

Median fixing time:

The fixing time is time taken by developer D to handle bug report BR. This factor is not directly given in a bug report. Thus, it is estimated for each record by the difference between the open and closed dates as shown in Equation 4. the score of developers in fixing time is calculated using Equation 5. Thus, fixing time score for a developer in topic K is the median value (fix time) of bug reports of topic K that solved by developer D. In fact, the less fixing time spent, the more experienced the developer will be.

$$time\ spent = closed\ date - open\ date \quad (4)$$

$$S(D)_{med\_time} = \tilde{X}\ (time\ spent\ in\ BR_D^K) \quad (5)$$

Developer score

The three factors participate in calculating the developer score. However, their importance is different from one another. For this reason, $\alpha_1, \alpha_2, \alpha_3$ are used to represent the effect of each factor in the developer score, as shown in Equation 6. The summation of the three weights must be equal to 1 and the value range of each of them must be greater than 0 and less than 1.

$$s(D) = \alpha_1 * S(D)_{sev} + \alpha_2 * S(D)_{com} + \alpha_3 * S(D)_{med\_time} \quad (6)$$

Bipartite graph: is used to represent the two sets (S1 is bug reports and S2 is developers) in form of vertices. Each element in the two sets is represented as a node in the graph. For each vertex in S1, there is a connection to each node in the second set. This link represents the cost (the score of the developer in the topic of the corresponding bug report). This type of graph is basically used for assignment problem in order to find the optimal correspondence to apply the objective function, whether to maximize or minimize the edge dependency. The purpose of using the bipartite graph is to find the subset of edges, such that no node in a set connect more than one edge. Bipartite graph is a powerful technique to simulate machine learning and data mining problems, such as recommendation systems, which are best described by bipartite graph, where the items are S1 and the users are S2. Also, the citation network analysis problems are simulated by bipartite graph, where vertices simulate the publications in both sides. Additionally, the question answer problems are also best described by bipartite graph [31]. Fig 3 shows how Bipartite graph is used to represent the triaging problem.

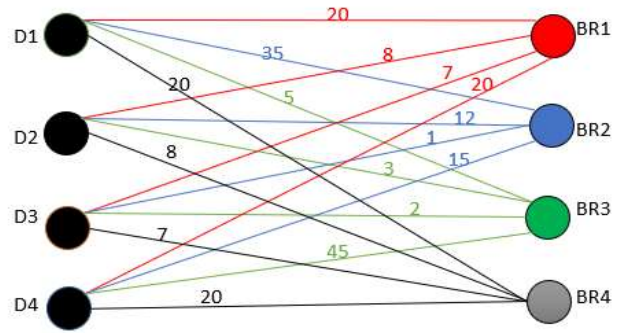

Fig. 2. Bipartite graph representing new bug reports and developers.

*6) Weights optimization phase*

The purpose of this phase is to find the best vector of weights $\alpha_1, \alpha_2, \alpha_3$, that can reduce the triaging cost and reach maximum time reduction.

Differential evolution algorithm (DE)

It is an optimization evolutionary algorithm for nonlinear problems, which depends on a stochastic approach. It was proposed by Storn and Price in 1997. The main steps of DE are shown below [32]. In initialization step, a random vector of weights is generated, then, using the objective function, this vector is improving iteratively, until reaching the best vector of weights, which makes the maximum fix time reduction.

1. NLTK :: Natural Language Toolkit

2. glove · PyPI
                                                                6

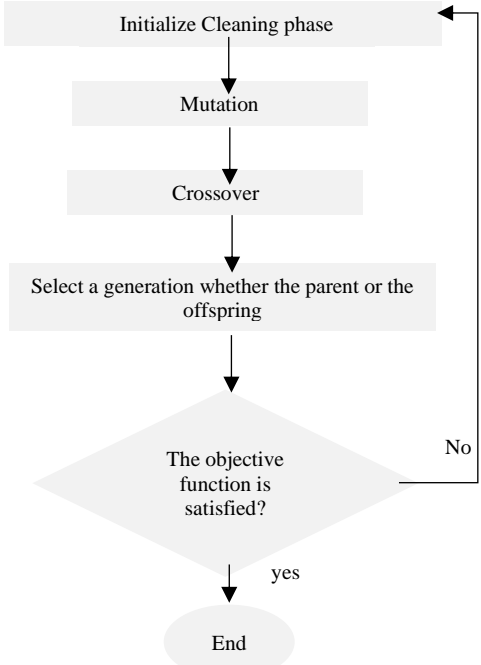

Fig. 3. Diffrencial evolution algorithm steps

*7) Assignment phase*

The purpose of this phase is to find the optimal developer for each new bug report (testing dataset) according to the topic of the bug report (required skills to be fixed) and the developer scores.

Gale-Shapely (GS): is named for David Gale and Lloyd Shapley. They firstly proposed it for college- admission problem in addition to the stable matching dilemma. Its input should be two types of participants that should be equal in size. The first input represents how members of the first set prefer the members of the second set to be matched to, the second set simulates the preferences for whom to be matched to the members on the first set. These two inputs are usually represented in matrix form. In brief, GS takes a number of rounds to be executed, each one proceeds in two stages. The first stage is about proposing each unmatched member in the first set proposes to its preferred member in the second set. Then, in the second stage, each agent in the second set accepts only one proposal among what have been proposed by the members in the first set in the previous stage. The algorithm goes to the end when all elements are matched.

TABLE I.  GALE-SHAPELY ALGORITHM

| | Gale-Shapely Algorithm |
|---|---|
| 1: | Declare the availability of each bug reports and developers. |
| 2: | **While** the bug report *BR* is available **do** |
| 3: |     *D*: = the first developer on the bug report *BR*'s preference list to whom the bug report *BR* has not yet assigned |
| 4: |     **If** *D* is available **then** |
| 5: |         The developer *D* is assigned to the bug report *BR* |
| 6: |     **Else** |
| 7: |         **if** *D* prefers *BR* to her assigned *BR*' **then** |
| 8: |             Re-assign *D* to *BR*; and re-declare the availability of *BR'* |
| 9: |         **Else** |
| |             *D* rejects *BR* |
| |         **end if** |
| |     **end if** |
| | **end while** |
| | **the assignments of the** *n* **couples are declared** |

To apply stable matching using Gale-Shapely algorithm and ensure that each bug report is assigned to a developer, the two number of columns must be equal to the number of rows in each of the input matrices, meaning that the number of bug reports must be equal to the number of developers. However, in triaging process, the number of bug reports is much bigger than the number of developers.

To overcome this dilemma, this paper extends the Gale-Shapely algorithm to be applied in an iterative manner. The testing dataset is split into small subsets, each one is of size that equals to the number of developers. Thus, the number of iterations to apply the Gale-Shapely will equal to the size of the testing dataset divided by the size of the developer set. In the last group in the testing dataset, that could be less than the number of developers, only a subset of the developers set will be considered. Fig 5 shows how the testing dataset is divided into subsets in applying iterative Gale-Shapely algorithm.

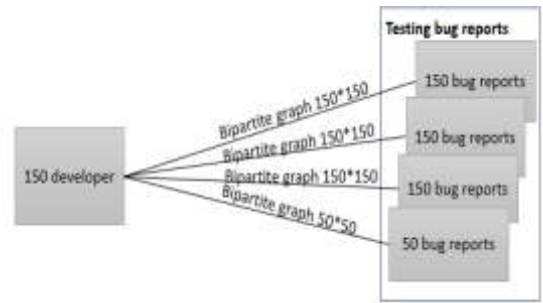

Fig. 4. The split of the new bug reports to fit the Gale-Shapely input requirements.

*B. Experiments:*

    *a) Datasets:*

In this paper Park's model is used as a reference to compere the proposed solution with. For this reason, the same dates filter has been used to extract bug reports from different bug repositories. The proposed model is evaluated on only resolved and closed bugs, to ensure the validity of dataset records. Only three datasets are used from Park's paper, Linux, Eclipse and Apache. However, Mozilla is excluded because in the specified dates, there is no records matched our query (bug reports in resolved and closed status). The actual time spent in fixing a bug is not recorded in the submitted bug reports. However, the open and closed dates are mentioned. Thus, an approximation for the fixing time is calculated by the difference between the two dates. A similar calculation used by Shihab [33] and Park [22].

---

1. NLTK :: Natural Language Toolkit
2. glove · PyPI

TABLE II. BUG REPORTS AND DEVELOPERS SETS SIZE AFTER AND BEFORE THE CLEANING PHASE

| TABLE | #BR before cleaning | #D before cleaning | #BR after cleaning | #D before cleaning |
|---|---|---|---|---|
| Linux | 15028 | 632 | 8538 | 183 |
| Eclipse | 186898 | 2021 | 117458 | 1048 |
| Apache | 29771 | 19 | 26331 | 18 |

For the three datasets, the exported bug report fields are bug ID, assignee, closing date, opening date, description, severity and component. Linux dataset is specified to be between 2002-11-14 and 2010-01-16, and between 2001-10-11 and 2010-01-22 in Eclipse and between 2001-01-22 and 2009-02-09 in Apache. Similar to filtering phase in Park's model, bug reports with assigned developer who fixed less than 10 bugs are removed. Some restrictions are added in this work, records with empty cell are removed and only bug reports with "RESOLVED" and "CLOSED" status are considered. Then, the records are stored in a chronological order. Therefore, the dataset can be split into training and testing. Because the testing part should be new with respect to the training section (historical dataset). Thus, the testing dataset is the later part when splitting the datasets. Table II shows the number of bug reports after and before the cleaning phase.

In Park's model, 80% of the dataset is used for training and 20% testing. This is a traditional data split in most of classification problems. However, triaging problem is a special case. It considers time factor, as each bug report can be considered as a historical record for the newly submitted reports. For this reason, time series split approach is used to split the dataset into training and testing in order to show the effect of the proposed approach on reducing the fixing time and normalizing the developer workload. It is considered as a modified version from the 10-fold cross validation approach. Only time-based problems can apply it. The size of the testing dataset is fixed (10% of all the dataset), however, the size of the training dataset is steadily increase by 10% in each iteration. Fig 6 shows how the dataset is split in each iteration. Because this paper introduces a comparison between the proposed solution and Park's solution, the two splitting approaches are included.

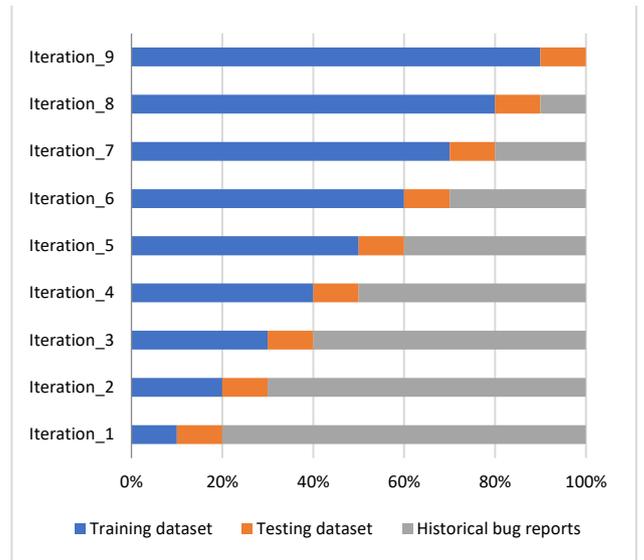

Fig. 5. Time series split technique

In the preprocessing phase, tokenization, stop word removal and lemmatization are applied. In this step, the description field is used as a representation for each bug (document). Then, the bug reports go into the labeling phase, which is the application of the topic modeling algorithm. To apply LDA algorithm, the number of topics should be given. Coherence score is one way to find the optimal number of K (topics), that should be given in applying the topic modeling. The expected result from the topic modeling is a vector of percentages simulating how each bug is related to each topic.

TABLE IV. EXAMPLE FOR LABELING EACH BUG REPORT BY A SINGLE TOPIC

|  | K1 | K2 | K3 | K4 | K5 | Kt | Label |
|---|---|---|---|---|---|---|---|
| BR1 | 50% | 3% | 10% | 2% | 15% | 28% | **K1** |
| BR2 | 20% | 20% | 3% | 40% | 11% | 6% | **K4** |
| BR3 | 7% | 80% | 1% | 1% | 2% | 9% | **K2** |
| BRn | 20% | 8% | 7% | 15% | 45% | 5% | **K5** |

As a labeling assumption, each bug report is labeled by the topic with the highest percentage (maximum relationship). The K range is set between 1 and 15. As an assumption, each document is labeled by the topic which takes the maximum percentage. In the next section, Fig 8 is a screenshot showing the optimal number of K. Table III shows an example for how each bug report is labeled.

TABLE V. DEVELOPER SCORE IN EACH TOPIC

|  | K1 | K2 | K3 | K4 | K5 | Kt |
|---|---|---|---|---|---|---|
| D1 | 20% | 30% | 5% | 35% | 5% | 5% |
| D2 | 8% | 8% | 9% | 12% | 3% | 60% |
| D3 | 7% | 80% | 1% | 1% | 2% | 9% |
| Dm | 20% | 8% | 7% | 15% | 45% | 5% |

The result of the topic modeling and the historical bug reports (training dataset) is used to set a score for each developer representing how each developer is skilled in each

1. [NLTK :: Natural Language Toolkit](#)

2. [glove · PyPI](#)   8

field (topic). Thus, the experience of each developer is represented by a vector. The length of the vector should be equal to the number of K. Each number in the vector is calculated by three main factors, bug severity, bug component and median value of the fixing time. Each of them is multiplied by a weight representing the importance of the factor in calculating the developer score. Table IV shows an example for setting a vector of scores for each developer.

The recommendation system is built using the resulted matrix by the application of topic modeling on testing dataset (row: bug reports, column: topics) and the transposed matrix of the scoring phase result (row: topics, columns: developers). Noted that the number of columns in the first matrix is just equal to the number of rows in the second one, the two matrices can be multiplied. The resulted matrix will be the recommendation of a set of developers for each bug report (row: bug report, column: developer). Equation 7 shows how the recommendation phase is calculated; $K\_assignment_M$ is the topic modeling result and $t(D\_Score_M)$ is the transposed developer scoring matrix.

$$DR = K\_assignment_M \times t(D\_Score_M) \qquad (7)$$

The transposed matrix of the recommendation system result (row: developers, column: bug report) will be the first input matrix to the Gale-Shapely algorithm. It shows as, how each developer is suitable for the bug reports in the testing dataset. To build the matrix which shows how each bug report in the testing dataset prefers the developers, the topic modeling of each bug report in the testing dataset is used. Accordingly, a matrix is built (rows: bug reports, columns: developers). Each row represents the score of all developers in the topic of the corresponding bug report. The term suitable means to what extent a couple of a bug report and a developer is a good match for each other according to the bug report topic (the skills required to fix the bug report) and the developer score (the experience of a developer in the corresponding bug report topic)

By applying the Gale-Shapely algorithm using the two previously mentioned matrices, a developer will be assigned to each bug report in one-to-one assignment. However, the number of bug reports should be equal to the number of developers, so the Gale-Shapely algorithm is applied iteratively. In each iteration a subset of the testing bug report is taken to be assigned to the developers set in one-to-one assignment.

*C. Evaluation matrices*

The proposed algorithm is evaluated against two main concepts, fixing time and developer working load.

The first aspect (fixing time) calculates how many days the assigned developers (proposed algorithm result) take to fix a set of bugs compared with the number of days that the real developers took to handle the same set.

$$TR = ((TS_{Real} - TS_{Assigned}) / TS_{Real}) * 100 \qquad (8)$$

In Equation 8 $TR$ simulates the percentage of time reduction. $TS_{Real}$ is the number of days that have been taken by the real developers to fix the bug reports in testing dataset. $TS_{Assigned}$ is the number of days that the resulted set of developers may take to fix the same set of bug reports. The value of $TR$ should be always positive and the higher it is, the more optimized the fixing time is, the more effective the proposed algorithm is.

The second aspect is the developer's workload, the distribution of the bug reports over the existing group of developers. In Equation 9, $R(RWL)$ is the developer workload range using manual triaging. It is calculated by the difference between the maximum and the minimum load. In Equation 10, $R(ResWL)$ is the developer workload range using the proposed algorithm. It should shrink with respect to the workload in the manual assignment.

$$R(RWL) = [\min(RWL) - \max(RWL)] \qquad (9)$$

$$R(ResWL) = [\min(ResWL) - \max(ResWL)] \qquad (10)$$

IV. RESULTS AND DISCUSSION

Experiments are done using two dataset split approaches. First, the time split approach. Second, Park's approach, 80% is training and 20% is testing. While the range of K set to be between 1 and 15.

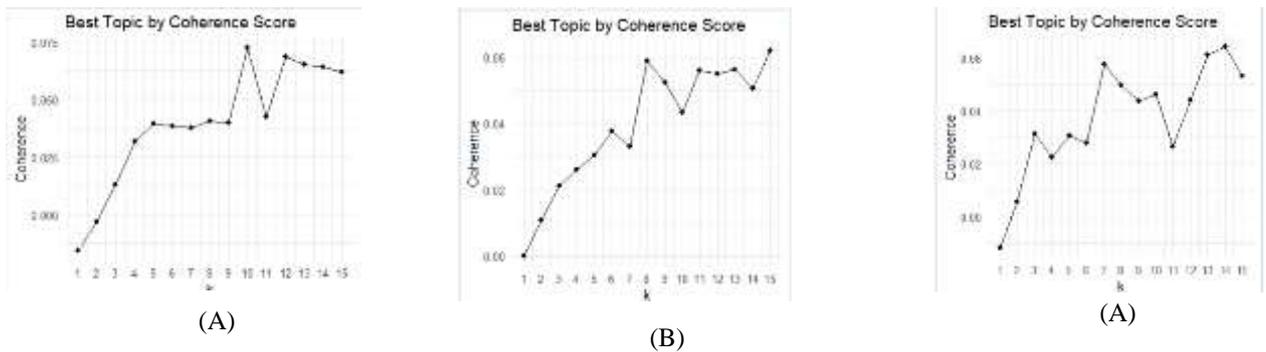

(A)  (B)  (A)

Fig. 6. Applying coherence score technique on (A) Linux dataset, (B) Eclipse dataset and (C) Apache dataset in the first iteration using time split approach.

1. NLTK :: Natural Language Toolkit

2. glove · PyPI

TABLE VI. THE EFFECT OF APPLYING THE ITERATIVE GALE-SHAPELY APPROACH OVER LINUX REPOSITORY IN TERMS OF THE WORK LOAD NORMALIZATION AND FIXING TIME REDUCTION, USING TIME SPLIT APPROACH.

| #I | #Ds | #TrD | #TsD | RWL | #K | Opt_W | ResWl | P_Acc (%) | T_Opt (%) |
|---|---|---|---|---|---|---|---|---|---|
| 1 | 1429 | 714 | 714 | 0-63 | 10 | 0.04424304 **0.6791318** 0.2766251 | 18-19 | 2.3 | 60.19 |
| 2 | 2208 | 1472 | 736 | 0-47 | 15 | 0.2261774 0.46163 **0.3121926** | 12-13 | 1.7 | 67.18 |
| 3 | 3067 | 2300 | 766 | 0-49 | 13 | 0.1645631 **0.7134467** 0.1219901 | 11-12 | 1.3 | 66.85 |
| 4 | 4007 | 3205 | 801 | 0-45 | 13 | 0.2997829 0.3199434 **0.3802737** | 9-10 | 0.62 | 71.68 |
| 5 | 4886 | 4071 | 814 | 0-39 | 14 | 0.1724038 **0.6559983** 0.1715978 | 8-9 | 1.3 | 79.63 |
| 6 | 5782 | 4956 | 826 | 0-50 | 14 | 0.1206084 0.1688999 **0.7104917** | 9-10 | 0.8 | 80.78 |
| 7 | 6673 | 5838 | 834 | 0-51 | 15 | 0.4485005 **0.4819387** 0.06956074 | 8-9 | 0.36 | 75.41 |
| 8 | 7610 | 6764 | 845 | 0-45 | 14 | **0.4369758** 0.1261696 0.4368546 | 8-9 | 0.47 | 81.04 |
| 9 | 8546 | 7691 | 854 | 0-47 | 15 | **0.4088339** 0.2421759 0.3489902 | 7-8 | 0.46 | 78.24 |

TABLE VII. THE EFFECT OF APPLYING THE ITERATIVE GALE-SHAPELY APPROACH OVER ECLIPSE REPOSITORY IN TERMS OF THE WORK LOAD NORMALIZATION AND FIXING TIME REDUCTION, USING TIME SPLIT APPROACH.

| #I | #Ds | #TrD | #TsD | RWL | #K | Opt_W | ResWl | P_Acc (%) | T_Opt (%) |
|---|---|---|---|---|---|---|---|---|---|
| 1 | 22731 | 11365 | 11365 | 0-523 | 15 | **0.5056276** 0.0402528 0.4541196 | 67-68 | 0.6 | 54.61 |
| 2 | 34341 | 22894 | 11447 | 0-283 | 14 | 0.188697 0.3240841 **0.4872189** | 40-41 | 0.2 | 50.23 |
| 3 | 46013 | 34509 | 11503 | 0-312 | 4 | **0.4506898** 0.1501899 0.3991203 | 33-34 | 0.2 | 46.95 |
| 4 | 57780 | 46224 | 11556 | 0-175 | 5 | 0.2046379 **0.5147875** 0.2805746 | 28-29 | 0.25 | 48.53 |
| 5 | 70004 | 58336 | 11667 | 0-214 | 4 | **0.4192756** 0.2750687 0.3056558 | 28-29 | 0.16 | 51.88 |
| 6 | 82202 | 70458 | 11743 | 0-232 | 4 | 0.1365474 0.1610199 **0.7024327** | 26-27 | 0.15 | 52.81 |
| 7 | 92982 | 81359 | 11622 | 0-539 | 4 | **0.4030455** 0.2742902 0.3226643 | 24-25 | 0.11 | 19.4 |
| 8 | 105681 | 93938 | 11742 | 0-245 | 4 | 0.3509971 **0.3618269** 0.287176 | 20-21 | 0.7 | 35.24 |
| 9 | 117458 | 105712 | 11745 | 0-544 | 3 | 0.0787019 0.2969453 **0.6243529** | 23-24 | 0.059 | 37.09 |

1. [NLTK :: Natural Language Toolkit](#)
2. [glove · PyPI](#)

TABLE VIII. THE EFFECT OF APPLYING THE ITERATIVE KUHN-MUNKRES APPROACH OVER APACHE REPOSITORY IN TERMS OF THE WORK LOAD NORMALIZATION AND FIXING TIME OPTIMIZATION, USING TIME SPLIT APPROACH.

| #I | #Ds | #TrD | #TsD | RWL | #K | Opt_W | ResWl | P_Acc (%) | T_Opt (%) |
|---|---|---|---|---|---|---|---|---|---|
| 1 | 5159 | 2579 | 2579 | 0-1257 | 14 | **0.5289167** 0.117633 0.3534503 | 214-215 | 7.52 | 49.26 |
| 2 | 7574 | 5049 | 2524 | 0 -1110 | 12 | 0.1787096 **0.5434278** 0.2778627 | 168-169 | 9.82 | 37.92 |
| 3 | 9990 | 7492 | 2497 | 0-1097 | 15 | **0.3638859** 0.3625812 0.2735329 | 156-157 | 8.21 | 44.19 |
| 4 | 12353 | 9882 | 2470 | 0-1088 | 15 | 0.22264 0.2791836 **0.4981764** | 154-155 | 7.41 | 99.74 |
| 5 | 15124 | 12603 | 2520 | 0-1109 | 13 | **0.4426502** 0.1557399 0.4016099 | 157-158 | 7.1 | 66.07 |
| 6 | 17922 | 15361 | 2560 | 0- 1128 | 15 | 0.3324834 0.1120948 **0.5554218** | 160-161 | 8.59 | 59.53 |
| 7 | 20725 | 18134 | 2590 | 0-1142 | 12 | 0.0810633 **0.5787595** 0.3401772 | 161-162 | 7.29 | 63.08 |
| 8 | 23528 | 20913 | 2614 | 0-1155 | 11 | 0.1447743 **0.4374856** 0.41774 | 163-164 | 7.07 | 66.28 |
| 9 | 26331 | 23697 | 2633 | 0-1164 | 13 | 0.4438633 0.0341686 **0.5219681** | 154-155 | 7.03 | 68.89 |

Fig 7 shows the results of applying the coherence score over the three repositories in the first iteration. The optimal number of topics in the first iteration in Linux is 10, in Eclipse is 15 and in Apache is 14.

TABLE VI, TABLE VII and TABLE VIII show the results of applying the iterative Gale-Shapely algorithm over the three projects, using time split approach, which applies 9 iterations for each project. #Ds is the total number of the dataset after cleaning. #TrD is the number of records in the training dataset. #TsD is the size of testing dataset. RWL is the min and max developer work load using the manual triaging. #K is the best number of topics that resulted from applying the coherence score on the training dataset in each iteration. Opt_W represents the optimal numbers for the three weights that used to calculate the developer score; severity, component and the median fixing time, respectively. ResWL is the normalized work load; as a result from applying the proposed algorithm. P_Acc is the prediction accuracy, however, this factor is out of this paper's scope. T_Opt is the main goal of this paper, it represents the optimized fix time.

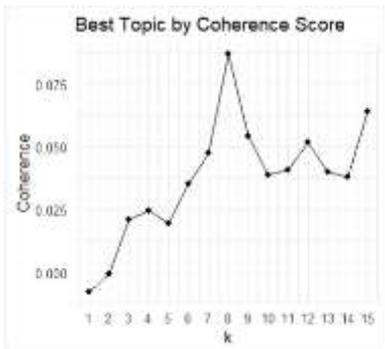
(A)

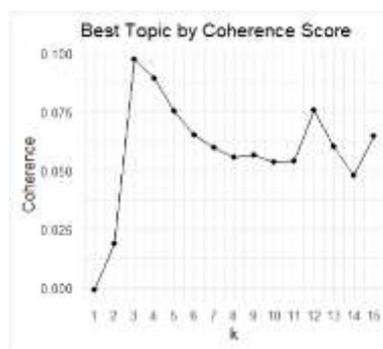
(B)

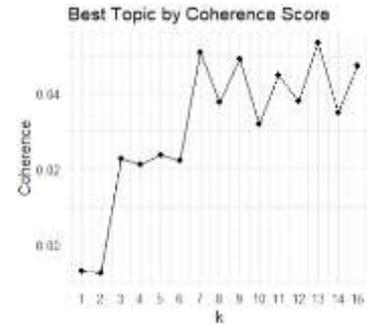
(C)

Fig. 7. Applying coherence score technique on (A) Linux repository, (B) Eclipse repository and (C) Apache repository.

1. NLTK :: Natural Language Toolkit
2. glove · PyPI

TABLE IX. THE EFFECT OF APPLYING THE ITERATIVE KUHN-MUNKRES APPROACH OVER LINUX, ECLIPSE AND APACHE REPOSITORIES IN TERMS OF THE WORK LOAD NORMALIZATION AND FIXING TIME OPTIMIZATION, COMPARED WITH COSTRIAGE APPROACH.

| Project | #Ds | #TrD | #TsD | RWL | #K | Opt_W | ResWl | P_Acc | T_Opt | R_CostTriage |
|---|---|---|---|---|---|---|---|---|---|---|
| Linux | 8538 | 6830 | 1707 | 0-94 | 8 | 0.3984924<br>0.1740531<br>**0.4274545** | 10-11 | 0.53 | **80.67 %** | 28.9 % |
| Eclipse | 117458 | 93966 | 23491 | 0-684 | 3 | **0.5056276**<br>0.0402528<br>0.4541196 | 34-35 | 0.07 | **23.61 %** | 10.6 % |
| Apache | 26331 | 21064 | 5266 | 0-2136 | 13 | 0.1738529<br>**0.5655792**<br>0.2605679 | 351-352 | 0.06 | **60.22 %** | 59.66% |

To allow comparison with Park's model, 80% training and 20% testing approach is also applied. Fig.8 shows the result of applying the coherence score method on the three datasets. The optimal number of K is 8, 3 and 13 in Linux, Eclipse and Apache respectively.

TABLE VIII shows the result of applying the iterative Gale-Shapely algorithm over the three repositories using 80% training and 20% testing approach. R_CostTriage is the result of applying Park's model on the same dataset using the same dataset split approach.

*A. Result analysis*

TABLE VI, TABLE VII and TABLE VIII show the results of applying the proposed algorithm on Linux, Eclipse and Apache repositories using time split approach. P_Acc value in the 9 iterations throughout the three tables is positive value. This means that using the proposed triaging model did an actual fix time reduction with different amounts of bug reports in each iteration across the three repositories. For example, results of Linux dataset in TABLE VI show that the optimal value of K ranges between 10 and 15. This indicates that there are variety of ideas throughout all bug reports. The highest time reduction is in iteration 8, where the fix time is optimized by 81.04%, compared with the manual triage. However, the first iteration recorded the least fix time reduction, 60.19%. By taking the average across all the 9 iterations, the fixing time is reduced by 73.44% of the original fixing time by the manual triage in Linux experiment, 44.08 % in Eclipse and 61.66 % in Apache.

On top of that, the range (min and max) workload of the developers is quite normalized in the three repositories, which means that applying the proposed model makes all developers work equally. This also leads to fast addressing to bug reports in the queue, which positively affects the total fixing time. From the financial perspective, the working hours for each developer will be lowered and accordingly salaries will be reduced.

Differential evolution algorithm plays a crucial role in adjusting the impact of the three factors in calculating the developer score. Results in TABLE IV, TABLE IIV and TABLE IIIV indicates that there is no pattern for $\alpha_1$, $\alpha_2$ and $\alpha_3$ values. So, it depends on the dataset records. For example, in Apache project results in TABLE IIIV, severity factor has the highest weight in calculating the developer score in iterations 1, 3 and 5, however component factor is the most important one in iterations 2, 7 and 8, and median fixing time is the essential one in iterations 4,6 and 9. In conclusion, the three factors are important. However, in each iteration, one of them is more important than the other two.

In Fig 8, the small number of K in Eclipse in diagram (B) indicates how related the data is, because a small number of topics can be a good representation for the whole dataset. However, in Eclipse project in diagram (C), the description field of the dataset include many aspects, that's why the number of topics is relatively larger. Thus, the description filed in Eclipse is more related to each other than in the Linux and Apache.

TABLE IX shows how the iterative Gale-Shapely algorithm strengthened by matrix factorization and differential evolution algorithm recorded outstanding optimization percentage in terms of the total bug fixing time, compared with Park's model. For example, in Linux, the proposed algorithm shows that 0.3984924, 0.1740531 and 0.4274545 are the optimal values for $\alpha_1$, $\alpha_2$ and $\alpha_3$, respectively. This means that approximated median fix time factor has the maximum importance in calculating the developer score, followed by severity and component factors. Using these three weights in calculating the developer score can reduce the fix time by over 80%, compared with the manual triage. This result is much better that the optimization percentage achieved by Park's model, which is almost 29%-time reduction. In addition, experiments on the other two projects, Eclipse and Apache also testify that the proposed algorithm is much more effective than Park's model.

In terms of the normalized developer work load, this factor is totally ignored in the manual triage. Also, it is not considered in Park's paper. However, the proposed algorithm achieved good results in equalizing the developer load. For example, in Eclipse project the range of RWL is [0-684]. This means that there are some developers, who are not assigned to any bug report, zero work load. Whereas, some other developers have heavy work load, assigned to 684 bug reports. In this case, some developers will be overwhelmed by many bug reports, while, other developers will be totally free. The proposed algorithm achieved good results in addressing this dilemma. Using iterative Gale-Shapely algorithm, one to one assignment, the work load will be normalized among the developers set, so, each developer in Eclipse repository will be assigned to 34 or 35 bug reports.

Considering that prediction accuracy is a minor goal for this paper, P_Acc value in the three experiments is very low. For example, in Linux repository P_Acc is 0.53%, this value means that the proposed algorithm assignment is similar to the manual assignment by 0.53%, which indicates that the manual triage requires 99.47% modifications in the assignment process to achieve 28.9% fix time optimization besides developer work load normalization.

---

1. [NLTK :: Natural Language Toolkit](#)
2. [glove · PyPI](#)

## V. THREATS TO VALIDITY

Certainly, there are few threats to ensure the validity of the proposed model:

- The proposed algorithm used as many bug fields as possible to ensure generalization in such an optimized and automated bug triaging optimization model. However, there are many other bug attributes may be included in the solution, such as the number of comments, reporter name, bug priority and the textual information in the comments section between the reporter and the bug fixer.

- The proposed solution is not a classification model, it is an optimization one instead. Thus, the prediction aspect is not highly considered in this paper. However, optimization techniques and classification algorithms could be combined to ensure the time reduction along with the high prediction accuracy.

- Because of the use of the differential evolution iteratively, the running time is relatively long especially in large datasets.

- For the comparison purpose, the dates used to filter the dataset in Park's model is used in this work. However, the cleaning phase is executed to our best of understanding, because minor information is mentioned in Park [22] about the cleaning phase.

- In real life scenario, some developers leave the project and others join, in different periods. However, the proposed model assumes that no changes happened in the developers set.

## VI. CONCLUSION AND FUTURE WORK

In conclusion, triaging bug reports consumes time, human resources and financial resources when using the manual approach, which puts huge load on only a certain group of developers. Therefore, some developers will be stressed and the others will have no work most of the time. Thus, the triaged bug reports will stay not addressed for long time because the developers are busy with other bug reports in the queue. For this reason, it can be concluded that the distribution of the bug reports on developers has a great effect on reducing the fixing time. Unlike the manual assignment, this automatic approach is designed, implemented and tested with intension to optimize the fixing time and equalize the developer workload. Gale-Shapely assignment algorithm supported by matrix factorization and differential evolution algorithm is used iteratively to automate the triaging process and assign a developer to each bug report in a way that optimizes the total fixing time besides normalizing the workload among developers.

As a future work, more assignment algorithms will be investigated to establish a comparative study between the effects of different algorithms on reducing the fixing time. In addition, more bug report attributes can be involved to ensure generalization. Also, reducing the execution time and considering the scenario of some developers can join or leave the project

## REFERENCES

1. NLTK :: Natural Language Toolkit

2. glove · PyPI

1. NLTK :: Natural Language Toolkit

2. glove · PyPI